\documentclass[%
 reprint,
superscriptaddress,
 amsmath,amssymb,
 aps,
prb,
]{revtex4-2}

\usepackage{graphicx}
\usepackage{epsfig}
\usepackage{dcolumn}
\usepackage{bm}
\usepackage{url}
\usepackage{hyperref}
\usepackage[dvipsnames]{xcolor}
\usepackage{siunitx}

\newcommand{\kappaph}{\kappa_\mathrm{ph}}

\newcommand{\tauph}{\tau_\mathrm{ph}(\omega)}

\begin{document}

\preprint{APS/123-QED}

\title{Unveiling the Thermoelectric Properties of Group III-Nitride Biphenylene Networks}

\author{G{\"o}zde {\"O}zbal Sargın}
\affiliation{Department of Fundamental Sciences, Air NCO Vocational HE School, Turkish National Defense University, Izmir, 35415 Turkey.}
\affiliation{Faculty of Engineering and Natural Sciences, Sabanci University, Istanbul, 34956 Turkey.}

\author{Kai Gong}
\affiliation{Department of Civil and Environmental Engineering, Rice University, Texas, 77005 USA.}
\affiliation{Rice Advanced Materials Institute, Rice University, Texas, 77005 USA.}
\affiliation{Ken Kennedy Institute, Rice University, Texas, 77005 USA}

\author{V. Ongun {\"O}z{\c{c}}elik} \email{ongun.ozcelik@sabanciuniv.edu}
\affiliation{Faculty of Engineering and Natural Sciences, Sabanci University, Istanbul, 34956 Turkey.}
\affiliation{Materials Science and Nano Engineering Program, Sabanci University, Istanbul, 34956 Turkey.}

\date{\today}

\begin{abstract}

After the synthesis of the carbon biphenylene network (C-BPN), research has increasingly focused on adapting elements from other groups of the periodic table to this lattice structure. In this study, the direction-dependent electronic, thermal, and thermoelectric (TE) properties of semiconducting group-III (group-III $=$ B, Al, Ga, In) nitride biphenylene networks are investigated using the non-equilibrium Green’s function formalism in combination with first-principles calculations. Phonon spectra and  force field molecular dynamics (MD) simulations were used to asses the dynamically and thermally stable structures. At room temperature, the lowest phonon thermal conductance values are obtained for InN-BPN, with $\kappaph$ = 0.12 nW/K/nm and $\kappaph$ = 0.21 nW/K/nm along the armchair and zigzag directions, respectively. The nearly dispersionless valence-band region between the $\Gamma-X$ symmetry points causes a sharp increase in the $p-type$ electronic transmission, which significantly enhances the $p-type$ thermoelectric figure of merit, $zT$. Among the investigated group-III nitride BPNs, InN-BPN exhibits the best performance, with a $p-type$ $zT$ value of 2.33 in the zigzag direction at 800~Kelvin.
\end{abstract}

\maketitle


\section{Introduction}

Graphene, the pioneering carbon based two-dimensional (2D) material, has profoundly influenced materials research over the past two decades, owing to its unique properties revealed through both experimental and theoretical studies~\cite{Geim2007,doi:10.1126/science.1102896}. Most graphene derivatives were initially predicted theoretically, while only a subset has been realized experimentally using advanced synthesis techniques~\cite{PhysRevLett.106.155703,JIANG2017370,C8RA07074A,LIMA2025117235,doi:10.1073/pnas.1416591112,LARANJEIRA2024100321,Du2017,C2TC00006G}. Among the carbon-based materials reported in the literature, 45 are classified as exhibiting $sp^2$ hybridization, in which all carbon atoms are fully threefold coordinated~\cite{GIRAO2023611}. In the last decade, the carbon biphenylene network (C-BPN), a recently discovered 2D carbon allotrope, has emerged as an increasingly attractive subject of research. Although C-BPN had been proposed in earlier theoretical studies~\cite{tyutyulkov1997structure,hudspeth2010electronic,karaush2014dft}, its successful realization on a gold surface was achieved in 2021 by Fan \textit{et al.} through the HF-zipping technique~\cite{doi:10.1126/science.abg4509,https://doi.org/10.1002/anie.201707272,D2NH00528J}. Unlike graphene, which possesses a honeycomb lattice, C-BPN consists of periodically ordered square, hexagonal, and octagonal rings of $sp^2$-bonded carbon atoms~\cite{demirci2022hydrogenated}. Experimental measurements and simulated differential conductance of C-BPN have demonstrated that the band gap of armchair-type C-BPN nanoribbons rapidly decreases with increasing ribbon width and eventually closes at a critical width of 21 BPR, indicating that 2D C-BPN exhibits metallic behavior~\cite{doi:10.1126/science.abg4509}.

The experimental realization of C-BPN synthesis via a bottom-up approach has stimulated growing theoretical efforts aimed at incorporating elements from other groups into its non-benzenoid framework~\cite{demirci2022stability, OZDEMIR2025118280, abdullahi2023stability, laranjeira2024zns, gorkan2023can, gorkan2022functional}. In this context, the electronic, mechanical, and optical properties of the boron nitride analogue of C-BPN (BN-BPN) have been thoroughly investigated~\cite{D3CP00776F}. In addition, the adsorption behavior and storage capacity of BN-BPN toward various atoms and molecules have been extensively explored using first-principles methods~\cite{Santos2023,MOHAMMED2025112535,Jardan2025}. Hybrid BPN structures composed of C, B, and N have been shown to be promising candidates for thermoelectric applications and as high-performance anode materials for alkali-metal batteries~\cite{D4NR02754J,D4NR01386G}. Furthermore, strain-engineered BPN systems have demonstrated potential for hydrogen evolution reaction applications~\cite{YUAN2024159555}. Among group-III nitrides, the InN analogue of C-BPN exhibits a strain-tunable band gap under tensile deformation, rendering it attractive for electronic device applications~\cite{doi:10.1021/acsomega.4c03511}. Using density functional theory calculations, Lima et al. reported carrier mobilities alongside the thermal and dynamical stability of GaN- and AlN-based BPN structures~\cite{LOPESLIMA2023107183}. A common characteristic shared by these systems is their pronounced direction-dependent electronic, vibrational, and mechanical properties. First-principles studies further confirmed that group III–V biphenylene networks possess both dynamical and thermal stability~\cite{ABDULLAHI2024113103}. Moreover, the incorporation of alkali metals (Li, Na, K), superalkali NLi$_4$, and Ca atoms significantly enhances hydrogen adsorption capacity, positioning BP-BPN as a promising hydrogen storage material~\cite{DJEBABLIA202433,doi:10.1021/acsaem.5c00897,D4RA07271E}. BP-BPN has also emerged as a high-potential anode material for Li/Na-ion, Mg-ion, and Li/Na–S battery technologies~\cite{HAOUAM2024160096,ZAINUL2024112805}.

In addition to electronic, structural, and mechanical properties, it is also critical to explore their thermoelectric and carrier transport properties to understand whether they hold potential for applications such as energy conversion and nanoelectronic devices. A high electrical response of a material to a temperature gradient between two ends, expressed as a large $zT$ value, can be determined by the mutually interrelated Seebeck coefficient ($S$), electrical conductivity/conductance ($\sigma$/$G_e$), and thermal conductivity/conductance ($\kappaph$/$G_{ph}$). Here we note that, 2D materials emerged as strong TE candidates after the pioneering works~\cite{hicks1993} by Hicks and Dresselhaus, who revealed that low-dimensional materials can have high thermoelectric efficiency due to quantum confinement effects. While pristine 2D C-BPN suffers from metallic behavior which limits its TE efficiency, its engineered forms hold great potential~\cite{PhysRevMaterials.9.024003}. Furthermore, hybrid compositions of carbon, boron, and nitrogen in the biphenylene network result in significantly higher $zT$ values~\cite{D4NR02754J,PhysRevB.107.045422}. Previously, Kumar et.al showed that BN-BPN has a maximum $zT$ with a value of 0.68 at 900~K under p-type doping~\cite{D3CP03088A}.

Despite extensive research in revealing various properties of C-BPN counterparts, their thermoelectric properties remain insufficiently studied. In this paper, we perform a series of systematical analyses  on the thermoelectric properties of group-III nitride BPNs using the Landauer formalism combined with the first-principles calculations. We show that, certain group-III nitride BPNs are thermally and dynamically stable through the use of the \textit{ab-initio} and force-field MD simulations, supported with phonon dispersion calculations. Flat-like valence band character gives rise to abrupt change in transmission spectrum near the Fermi level, which results in notably large Seebeck coefficients. When incorporating the effect of lattice thermal conductance, remarkable p-type $zT$ value of 2.33 at 800~K is achieved for InN-BPN.

\section{Methods}
First-principles calculations were performed using the Vienna \textit{ab-initio} Simulation Package (VASP)~\cite{hafner1997vienna,kresse1996efficient} and the QuantumATK software~\cite{Smidstrup_2020}, based on density functional theory (DFT). The Perdew–Burke–Ernzerhof (PBE) exchange–correlation functional within the generalized gradient approximation (GGA) framework was employed for structural optimization and for the calculation of electronic, phononic, and transport properties, using a linear combination of atomic orbitals (LCAO) basis set. Norm-conserving pseudopotentials from the PseudoDojo database were employed to describe the core states, while a high-quality numerical atomic orbital basis set was used to expand the wavefunctions~\cite{VANSETTEN201839}. The density mesh cutoff was chosen as 75 Hartree, and the Brillouin zone was sampled using a 17$\times$10$\times$1 \textit{k}-point grid. For geometry relaxation, the force and stress tolerances were set to 10$^{-3}$ eV/$\SI{}{\angstrom}$ and 10$^{-4}$ eV/$\SI{}{\angstrom^3}$, respectively.

The dynamical matrix and phonon band structures were calculated with ultra basis set size and using the finite-difference method. A 5x3x1 supercell was employed to compute the second-order interatomic force constants. \textit{Ab-initio} molecular dynamics (AIMD) simulations are conducted to assess thermal stability utilizing a microcanonical ensemble at a constant temperature of 300~K. The total simulation time was 6 ps, with a time step of 1 fs. The stabilities of the BPN structures were further evaluated using force field MD simulations, facilitated by the QuantumATK NanoLab software package.~\cite{Smidstrup_2020} following the procedures similar to our previous study~\cite{PhysRevMaterials.9.024003}. Force field parameters for AlN,  BN, GaN and InN BPN structures  were adopted from previous studies~\cite{zhou2013molecular, moon2007theoretical, zhou2015towards}. Each model consists of a 10x20 supercell containing 2400 atoms. The systems were heated from 0 to 2000 K over 4 ns (0.5 K/ps) under periodic boundary conditions using the NVT ensemble, with temperature controlled by a Nose-Hoover thermostat~\cite{martyna1992nose}. Structural snapshots at representative temperatures prior to structural degradation (400 K for GaN and InN, and 800 K for BN) were extracted and subsequently equilibrated at the corresponding temperatures for 10 ns to assess thermal stability. A timestep of 1 fs was used for all simulations.

Here, we consider the group-III nitride BPNs as pristine 2D materials (free of defects, vacancies, or disorder) and calculate the electronic transmission spectra, $\tau(E)$,  by summing over the available conducting modes in the electronic band structure at each energy for both the armchair and zigzag directions. Number of k-points are 1$\times$101$\times$1 and 101$\times$1$\times$1 for the armchair and zigzag directions, respectively. Phonon transmission calculations are conducted based on the non-equilibrium Green's function method by following\cite{10.1063/1.3531573},
\begin{equation}
    \tauph = Tr[\Gamma_L G_c^R \Gamma_R G_c^A]
\end{equation}

After computing $\tau(E)$, electronic thermoelectric coefficients i.e. electrical conductance $G_e=e^2 L_0$, Seebeck coefficient (thermopower) $S=(L_1/L_0)/eT$, power factor $PF = S^2G$ and electrons contribution to the total thermal conductance $\kappa_{el}=(L_2-L_1^{2}/L_0)/T$ can be expressed via $L_n$ integrals as following within the Landauer formalism~\cite{sivan1986multichannel,esfarjani2006thermoelectric},
\begin{eqnarray}
L_n(\mu,T) = -\frac{2}{h}\int \tau(E) (E-\mu)^{n} \left(-\frac{\partial f_{FD}}{\partial E}\right) dE
\end{eqnarray}
where $\partial f_{FD}(E)/\partial E$ is the derivative of the Fermi-Dirac distribution function and $\tau(E)$ is the energy dependent transmission function. $\partial f_{FD}(E)/\partial E$ is also called fermi window function which describes energy of window over conducting channels that contribute to the electronic transport. 
Similarly, Landauer formalism is conducted to obtain phonon thermal conductance $\kappa_{ph}$ which is expressed as~\cite{datta1997electronic,rego1998quantized,sevinccli2019green},
\begin{eqnarray}
\kappa_{ph} = \frac{1}{2\pi} \int \tau_{ph}(\omega) \hbar \omega \left(\frac{\partial f_{BE} (\omega, T)}{\partial T}\right)d\omega
\label{kappaph_eq}
\end{eqnarray}
Here, $\tau(\omega_{ph})$ is phonon transmission spectrum per width which is performed using a $101\times101\times1$ q-point sampling.
Thermoelectric figure of merit $ZT=S^2GT/(\kappa_{el}+\kappa_{ph})$ can be predicted after obtaining electronic and phonon thermal transport parameters~\cite{data}.

\section{Results}

\subsection{Structural Properties}
\begin{figure}[h]
\includegraphics[width=\linewidth]{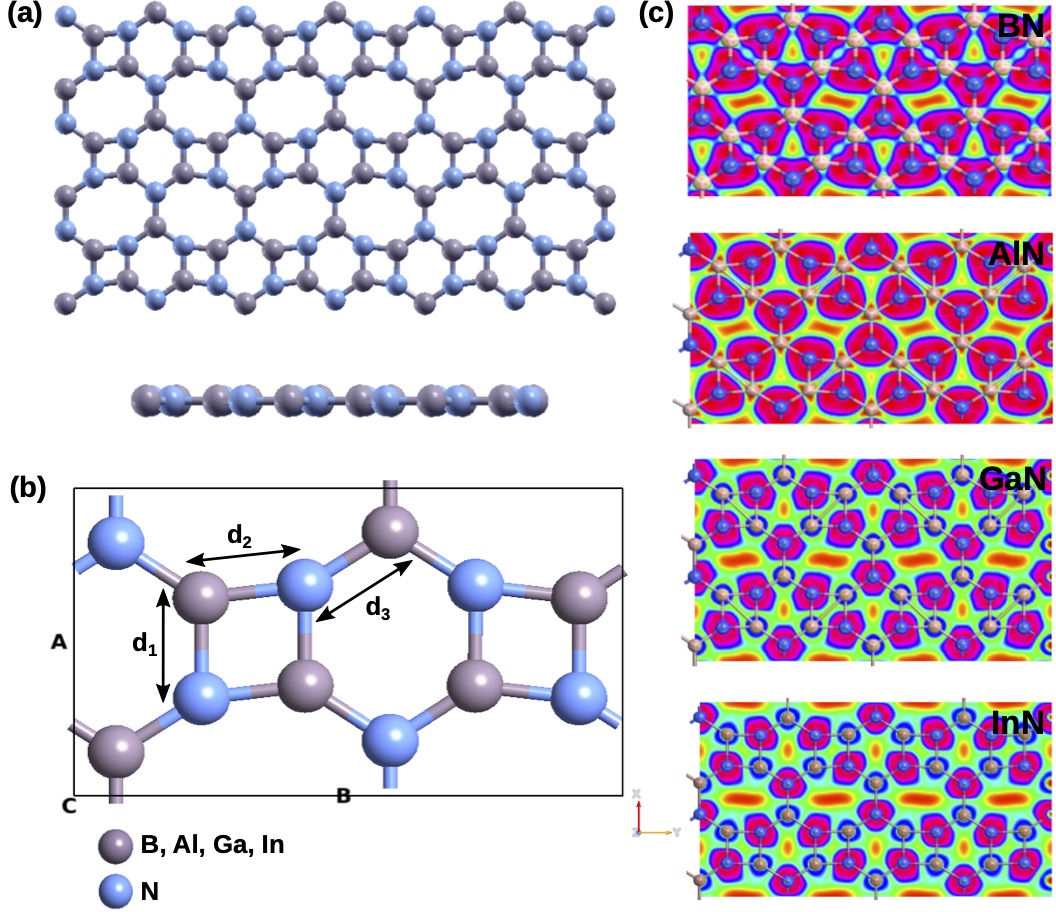}
\caption{\label{fig:fig1}(a) Prototype of the top and side view of the optimized crystal structure, (b) Unit cell of the structures and bond lengths between atoms (c) Electron localization function (ELF) of BN, AlN, GaN and InN biphenylenes, respectively.}
\end{figure}

\begin{figure}[htbp]
\includegraphics[width=\linewidth]{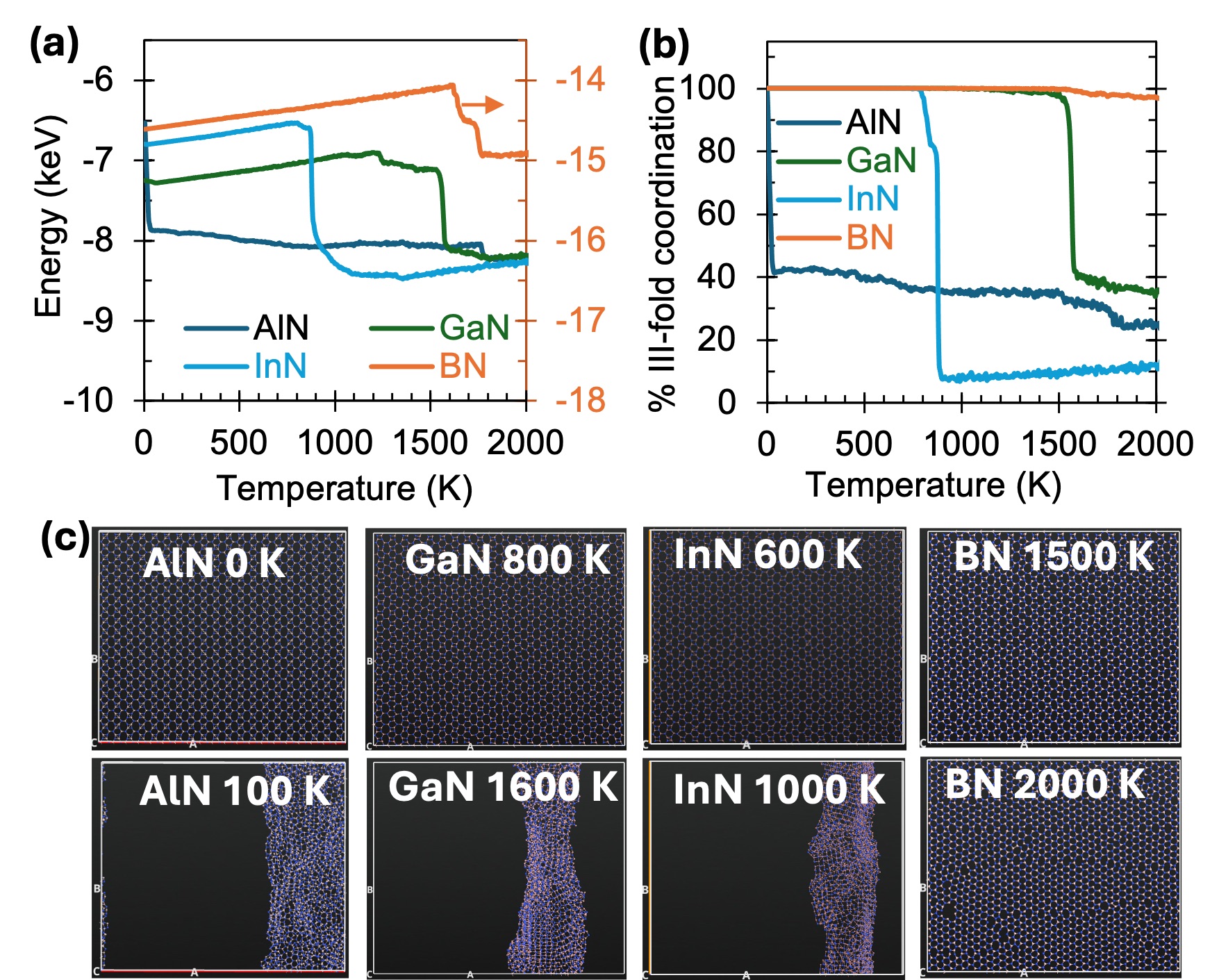}
\caption{\label{fig:figmd1} Evolution of (a) total energy and (b) the fraction of III-fold coordination (i.e., the number of Al, Ga, In, or B atoms surrounding N within its first coordination shell) in AlN, GaN, InN, and BN as the temperature increases from 1 to 2000 K over 6 ns. (c) Representative structural snapshots of AlN, GaN, InN, and BN at selected temperatures during the heating process. }
\end{figure}

Unit cell of the optimized geometry depicted in Fig.~\ref{fig:fig1}  consists of 12 atoms, 6 of which are group-III atoms and 6 of which are N atoms. Although slight distortions occur in the square, hexagonal, and octagonal rings, the planar geometry is preserved when compared to carbon biphenylene. High-temperature stability tests using force field MD were used to further justify the structural stabilities at elevated temperatures. Fig.~\ref{fig:figmd1} shows the temperature driven evolution of both energetic and local structural order in AlN, GaN, InN, and BN biphenylene structures during heating from 1 to 2000 K over a 6 ns MD trajectory. The monotonic increase in total energy shown in Fig.~\ref{fig:figmd1}(a) reflects the progressive thermal excitation of the systems, while deviations in slope and relative magnitudes highlight material specific responses to heating. More critically,  Fig.~\ref{fig:figmd1}(b) tracks the fraction of III-fold coordination around nitrogen atoms (defined by the number of Al, Ga, In, or B neighbors within the first coordination shell) serving as a sensitive descriptor of bond stability and local structural integrity. The gradual reduction in the III-fold coordination with increasing temperature indicates the onset of bond breaking and coordination loss. The rate and extent of this loss differs among the materials, thereby revealing their relative thermal robustness. These trends are further corroborated by the representative atomic configurations shown in  Fig.~\ref{fig:figmd1}(c), where the progression from ordered networks at low temperatures to increasingly disordered, partially amorphous structures at elevated temperatures is clearly observed, particularly for the less thermally stable nitrides.

\begin{figure}[htbp]
\includegraphics[width=\linewidth]{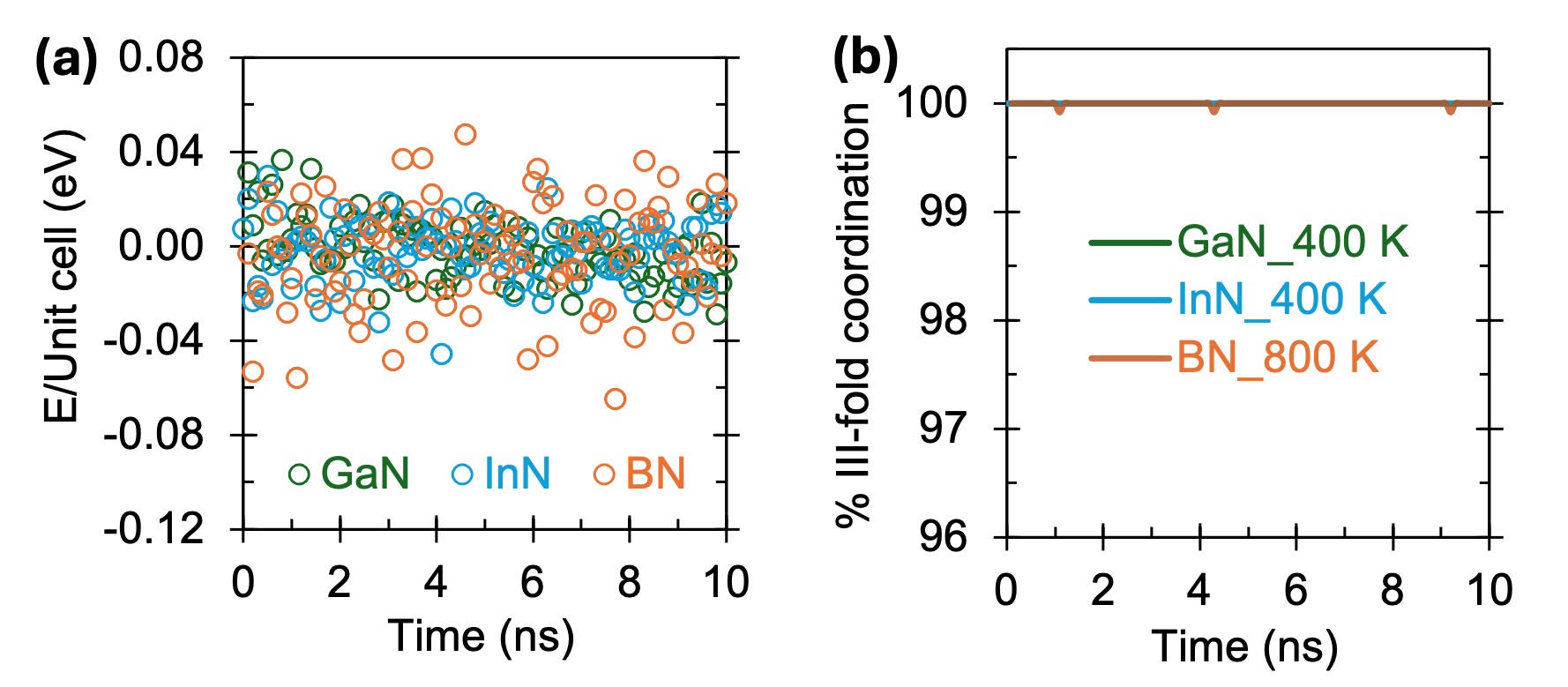}
\caption{\label{fig:figmd2} (a) Fluctuations in total energy (eV per unit cell) and (b) the fraction of III-fold coordination of N in GaN, InN, and BN equilibrated at 400 K, 400 K, and 800 K, respectively, over a 10 ns simulation. }
\end{figure}

Similarly, Fig.~\ref{fig:figmd2} examines the energetic stability and local coordination dynamics of GaN, InN, and BN  biphenylene structures under equilibrated thermal conditions. Fig.~\ref{fig:figmd2}(a) shows the temporal fluctuations of the total energy (eV per unit cell) during 10 ns simulations performed at 400 K for GaN and InN biphenylene and at 800 K for BN biphenylene, where the relatively narrow energy fluctuations indicate that each system has reached and maintained thermal equilibrium. Fig.~\ref{fig:figmd2}(b) presents the corresponding evolution of the fraction of III-fold coordination around nitrogen atoms, providing a quantitative measure of the stability of the local bonding environment under sustained thermal excitation. The persistence of near-100\% III-fold coordination over the simulation time demonstrates the robustness of the nitride networks at these temperatures, while subtle differences in fluctuation amplitude reflect material-dependent coordination dynamics and intrinsic thermal resilience.

\subsection{ Electronic Properties}

Group-III nitride BPNs have $P_{mma}$ space group (No.51) and $D_{2h}(mmm)$ point group. The high symmetry paths to be followed in computing the electronic and phonon bands in the  Brillouin zone of the unit cell is shown in Fig.~\ref{fig:fig1} (b).
As listed in Table 1, lattice parameters of the optimized structures are in order as follows \( \text{BN} < \text{AlN} < \text{GaN} < \text{InN} \), which are also in line with a previous study that used a plane wave basis set~\cite{demirci2022stability}. Bond lengths between the group-III elements and nitrogen are increasing with the enhanced lattice vectors of the compounds.

\begin{figure}[htbp]
\includegraphics[width=\linewidth]{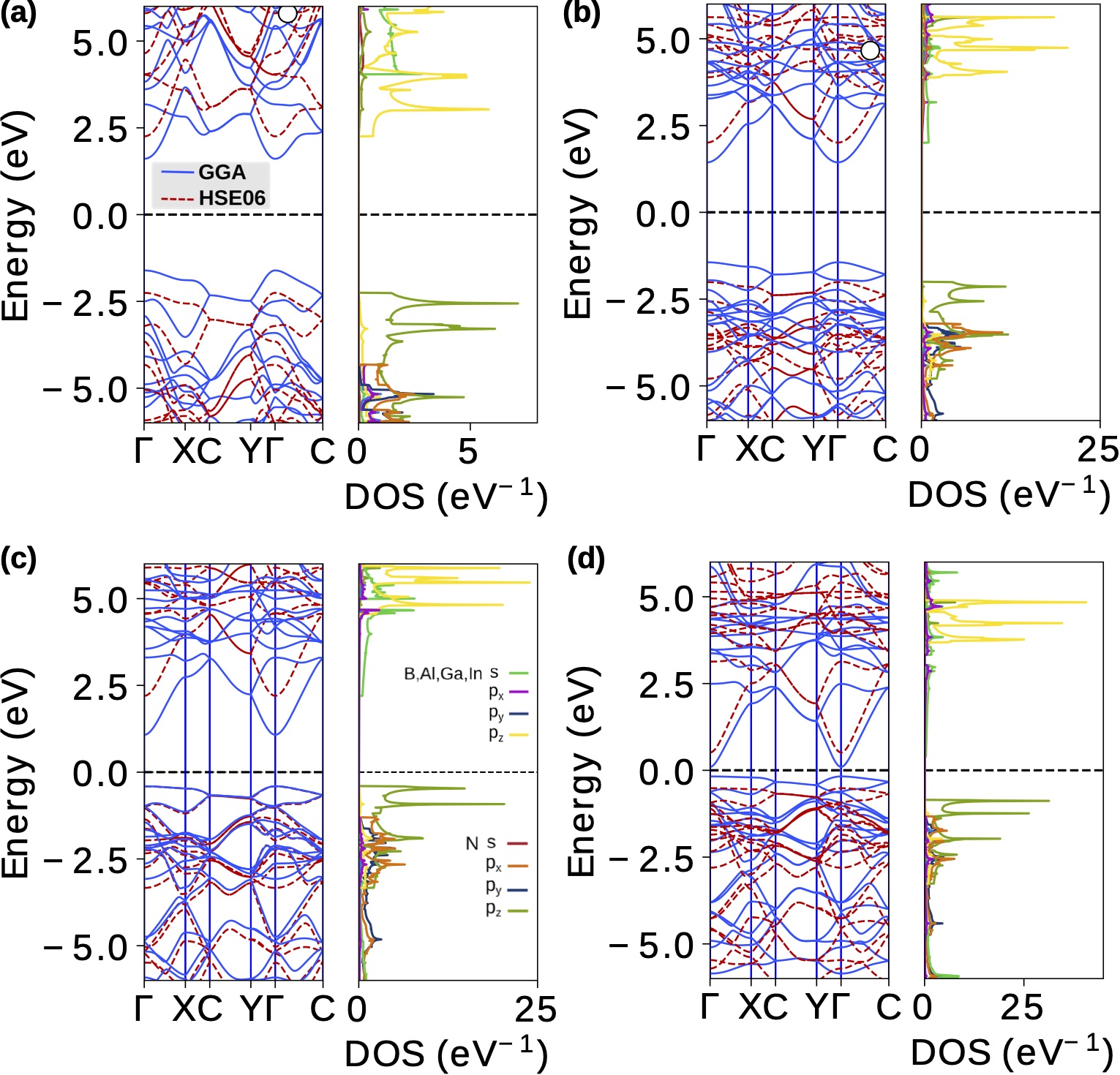}
\caption{\label{fig:fig2}Calculated electronic band structures based on PBE and PBE+HSE06 functionals and obtained projected density of states using  PBE+HSE06 functionals of group-III nitride biphenylenes. }
\end{figure}

Fig. \ref{fig:fig1} (c) demonstrates the electron localization function (ELF), which provides information on the bonding behavior of group-III nitride BPNs. The ELF map utilizes a color scale to represent electron localization: dark pink for fully localized electrons (0.97), light blue for partially delocalized ones (0.49), and red for regions with low electron presence (0.0). According to the ELF map, covalent bonding between group-III elements and nitrogen is weakened, which can be attributed to the enlarged lattice vectors and bond lengths. 
The electronic band structures of group-III nitride BPNs based on GGA and GGA+HSE06 functional are demonstrated in Fig.~\ref{fig:fig2}. BN BPN exhibits the largest band gap with a value of 3.22 eV (4.51 eV) while InN BPN has the smallest band gap with a value of 0.11 eV (1.13 eV). For all the structures, the $p_z$ orbital of the nitrogen atom plays a dominant role in forming the valence band. The conduction band of AlN, GaN, and InN biphenylene structures originates from the 
$s$ orbitals, while in BN biphenylene, the conduction band is primarily formed by the contribution of the $p_z$ orbitals of the boron atoms. Site projected DOS reveals that B, Al, Ga, and In, which are positioned at the corners of the square, hexagon, and octagon, give the dominant contribution to the conduction band edge. Similarly, nitrogens which are neighbours of the three polygons are the main contributors to the electronic states at the VBM. 

Since the CB edges of all structures are more dispersive than their VB edges, DOS takes smaller values. However, flat-like VB along $\Gamma$-X gives rise to a high DOS peak around the fermi level. The effective masses of charge carriers at the bands extrama are defined as 
\begin{equation}
  \left(\frac{1}{m^{*}}\right)_{ij}=\frac{1}{\hbar^2} \left(\frac{\partial E_n^2 (\mathbf{k})}{\partial k_i\partial k_j}\right),\hspace{0.4cm} i,j=x,y,z
\end{equation}

In order to increase the accuracy of the effective masses due to the non-parabolic band edges, finite difference method is used instead of quadratic band fitting. Considering the anisotropic nature of the structures, direction-dependent effective masses are listed in Table 1. The highest anisotropy in $m_e^*$ is observed for BN-BPN. For AlN, GaN and InN $m_e^*$ values along x- and y-directions are very close to each other. The conduction band edge becomes more dispersive and $m_e^*$ values decrease in heavier group-III elements. Contrarily, the hole effective masses $m_h^*$ are increasing with increasing atomic number and the topmost valence band exhibits almost non-dispersive character along the x-direction. Since, valence and conduction band dispersions are crucial for Seebeck coefficient, Fermi surfaces of the structures are calculated with a Fermi level shift 0.1 eV, as illustrated in Fig.~S1. Flat-like VB shape of structures along the x-direction are more pronounced in GaN and InN BPNs. Color gradients reflect the fermi velocity $\nu_F$ of holes and electrons. Moving from BN to InN, the variation in hole $\nu_F$ throughout the VB becomes less significant in parallel with increasing $m_h^*$.  

\begin{table*}[ht]
\centering
\label{tab:table1}
\caption{Lattice parameters, bond lengths, effective mass values of electrons and holes along armchair(x) and zigzag(y) direction, and electronic band gap values of group-III nitride BPNs. }
\begin{ruledtabular} 
\begin{tabular}{cccccccccccc} 
    \textbf{2D-BPN}&\textbf{$A/B$} & \textbf{$d_1$} & \textbf{$d_2$} & \textbf{$d_3$} & \textbf{$m^*_e$ (x/y)} &\textbf{$m^*_h$ (x/y)} & \textbf{$E_g^{GGA}$} & \textbf{$E_g^{HybridGGA}$} &\textbf{$E_g^{PBE+HSE06}$}&\\
    &($\SI{}{\angstrom}$)&($\SI{}{\angstrom}$)&($\SI{}{\angstrom}$)&($\SI{}{\angstrom}$)&(m$_e$)&(m$_e$)&(eV)&(eV)&(eV)&\\
    \hline
BN    &4.55/7.66 & 1.48 & 1.46 & 1.44 &0.81/0.37  &2.98/0.43 &3.22  &4.51&4.49\footnotemark[1]\\
AlN    &5.62/9.60 &1.85  & 1.81 &1.79  &0.56/0.52  &4.30/1.12& 2.88 &4.02 &3.53\footnotemark[1] \\
GaN    &5.90/9.96 & 1.93 & 1.89 & 1.86 &0.24/0.21  &8.18/0.93 &1.48  &2.58&2.65\footnotemark[1]  \\
InN    &6.59/11.12& 2.15 &2.10 &2.08  & 0.10/0.09 &16.15/1.10 &0.11&1.02  &1.13\footnotemark[1]  \\
\end{tabular}
\footnotetext[1]{\cite{demirci2022stability}}
\end{ruledtabular}
\end{table*}

\begin{figure}[h]
\includegraphics[width=\linewidth]{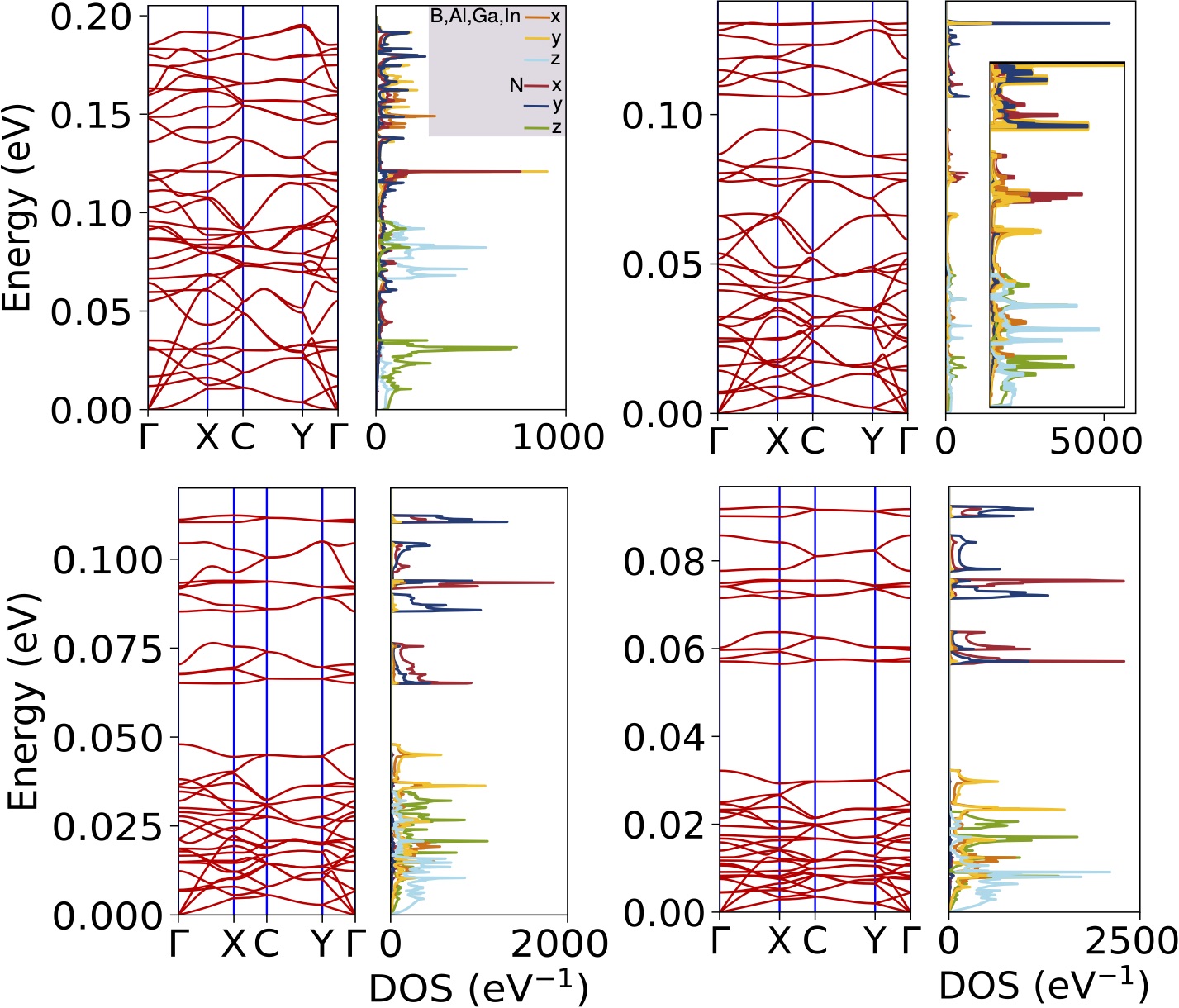}
\caption{\label{fig:fig3} Calculated phonon branches and projected density of states based on x, y and z movements of BN, AlN, GaN, InN biphenylenes, respectively.}
\end{figure}

\subsection{Vibrational Properties}
Phonon dispersions of BPNs have 36 phonon modes due to each unit cell of the biphenylene structures having 12 atoms. The dynamical stability of biphenylene structures was assessed by examining the phonon dispersion curves, as presented in Fig.~\ref{fig:fig3}. The lack of imaginary frequencies for all q-points in brillouin zone confirms the material's dynamical stability. From Boron to Indium, the highest phonon frequencies are reduced due to the increasing atomic mass and bond lengths of the compounds. BN-BPN has the maximum phonon frequency with a value of 0.197 eV (47.6 THz), while the heaviest compound InN-BPN exhibits highest frequency with a value of 0.092 eV (22.3 THz). All of the biphenylene structures exhibit a phonon band gap in their optical phonon spectrum except BN BPN. As the difference in atomic mass of the constituent atoms increases, the phonon band gaps enlarge between low-lying optical and high-range optical modes. The absence of accessible phonon transport channels between optical modes will tailor the lattice thermal conductance of the structures, especially at elevated temperatures. In addition, increasing the atomic mass of the compounds leads to flattened phonon bands, which signifies slower phonon propagation. In the lightest compound, BN biphenylene, the phonon modes vibrating in the z-direction of the nitrogen atom dominate the PhDOS at low frequencies. At the mid-range frequencies, PhDOS is primarily influenced by out-of-plane vibrations of boron atoms. Interestingly, a significant peak appears in the PhDOS, resulting from the combined effect of the degeneracy of the two optical phonon bands along the $C-Y$ direction and their flat dispersion.  The highest PhDOS peak has appeared in AlN-biphenylene due to the two ultra-flat highest optical phonon bands becoming degenerate. As the mass difference between group-III elements and nitrogen atoms increases, vibrations along the x- and y-directions of nitrogen lead to several peaks in PhDOS at high-range frequencies. Group III-nitride biphenylene structures possess $D_{2h}$ symmetry and space group of Pmma (No.51). 

\subsection{Transport Properties}

\subsubsection{Phonon Transport}

\begin{figure}[htbp]
\includegraphics[width=\linewidth]{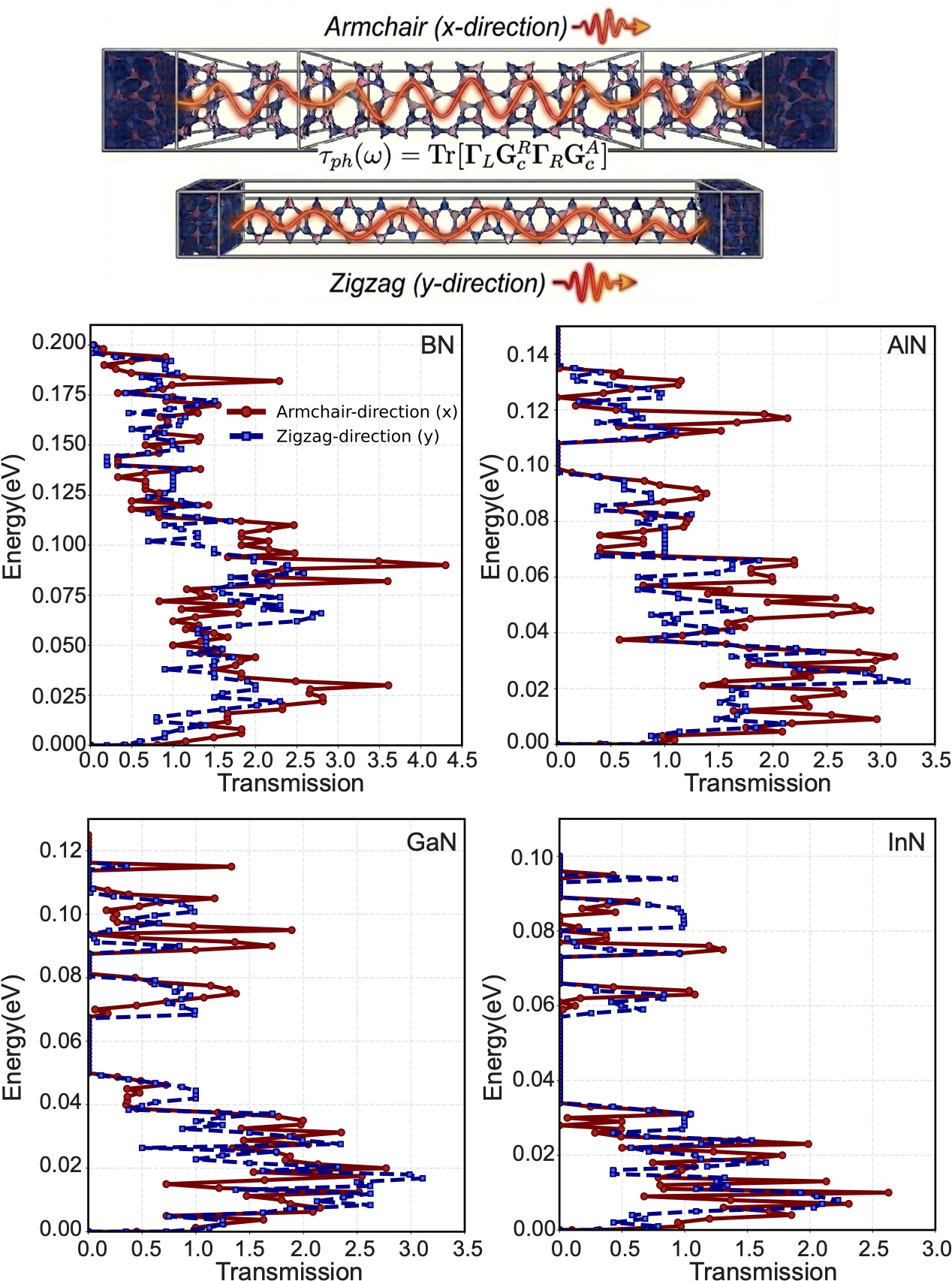}
\caption{\label{fig:fig4} The phonon transmission spectra of BN, AlN, GaN and InN BPN are shown for armchair and zigzag directions,. Device models are illustrated along the transport direction at the top.}
\end{figure}

\begin{figure}[htbp]
\includegraphics[width=\linewidth]{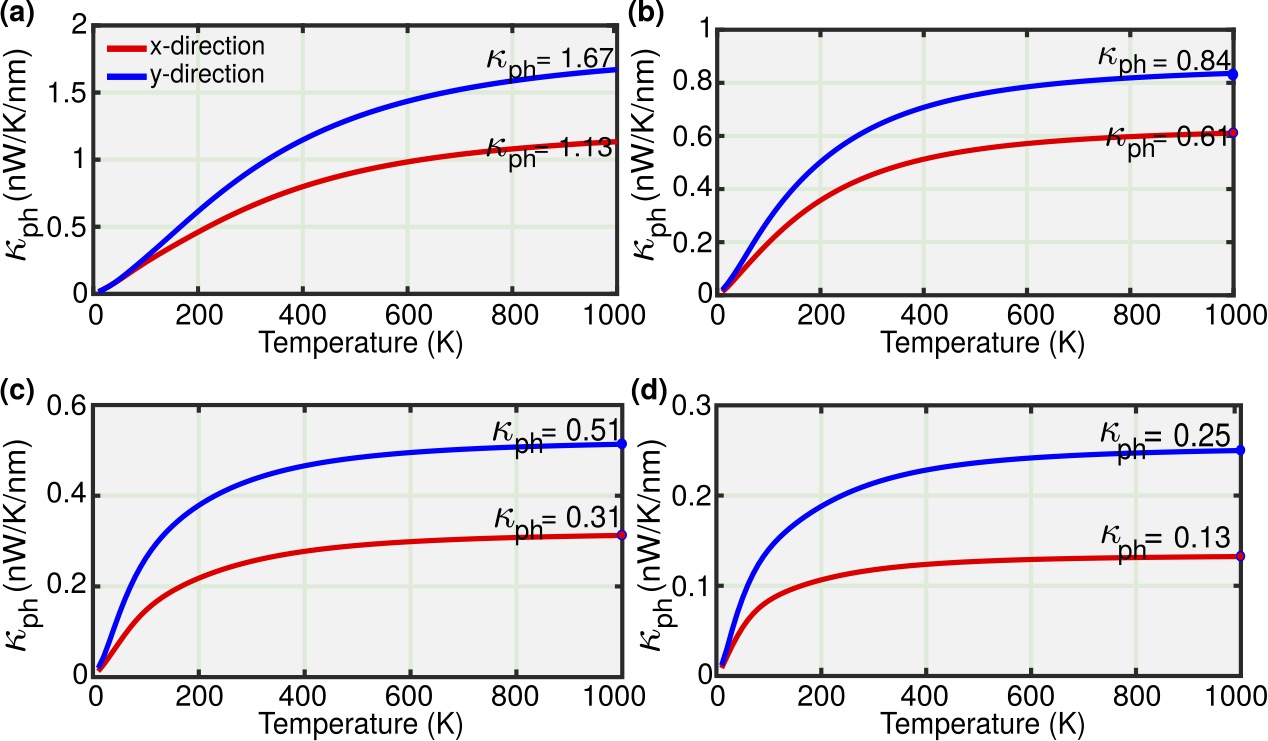}
\caption{\label{fig:fig5}Lattice thermal conductance behavior of group-III nitride biphenylenes for both armchair and zigzag direction up to 1000~K.}
\end{figure}

In order to obtain phonon transport properties within the ballistic framework, two-probe device model is used for armchair and zigzag directions of group-III nitride BPNs. As shown in Fig.~\ref{fig:fig4}, the nanoscale devices contain three regions, left electrode, right electrode, and central region, which are made of the same investigated BPN. For the armchair and zigzag systems, both electrodes are constructed using one unit cell, whereas the central region is built from five unit cells. 
In parallel with their phonon dispersion curves, phonon transmission spectrum of the lightest compound BN-BPN reaches the highest energy, whereas in InN-BPN the $\tauph$ is confined within a narrower energy interval. From AlN-BPN to InN-BPN phonon transmission gap increases due to the phonon band gaps between optical frequencies. The lack of available phonon transport channels in these gaps also causes $\tauph$ to drop to lower values. The strong anisotropy in $\tauph$ along the armchair and zigzag directions will result in corresponding anisotropic $\kappaph$.
Following the evaluation of the $\tauph$ of group-III nitride BPNs, their phonon thermal conductance behavior is determined using Eq~\ref{kappaph_eq}. Variation of the phonon thermal conductance as a function of temperature is illustrated in Fig.~\ref{fig:fig5}. The lightest BN-BPN shows the highest $\kappaph$ along both x- and y-directions. As the average atomic mass of the compound increases, $\kappaph$ values decrease for all temperatures. The heaviest InN-BPN exhibits the smallest $\kappaph$ with values of 0.13 $nW/K/nm$ and 0.25 $nW/K/nm$ along x- and y-directions, respectively. The strength of the anisotropy degree of heat transport $\kappaph^y / \kappaph^x$ of the BPN compounds is in order \( \text{InN} > \text{GaN} > \text{BN} > \text{AlN} \) at 1000~K. Based on $\kappaph$ values, it is evident that InN-BPN possesses a clear advantage in terms of thermoelectric efficiency.

\begin{figure}[htbp]
\includegraphics[width=\linewidth]{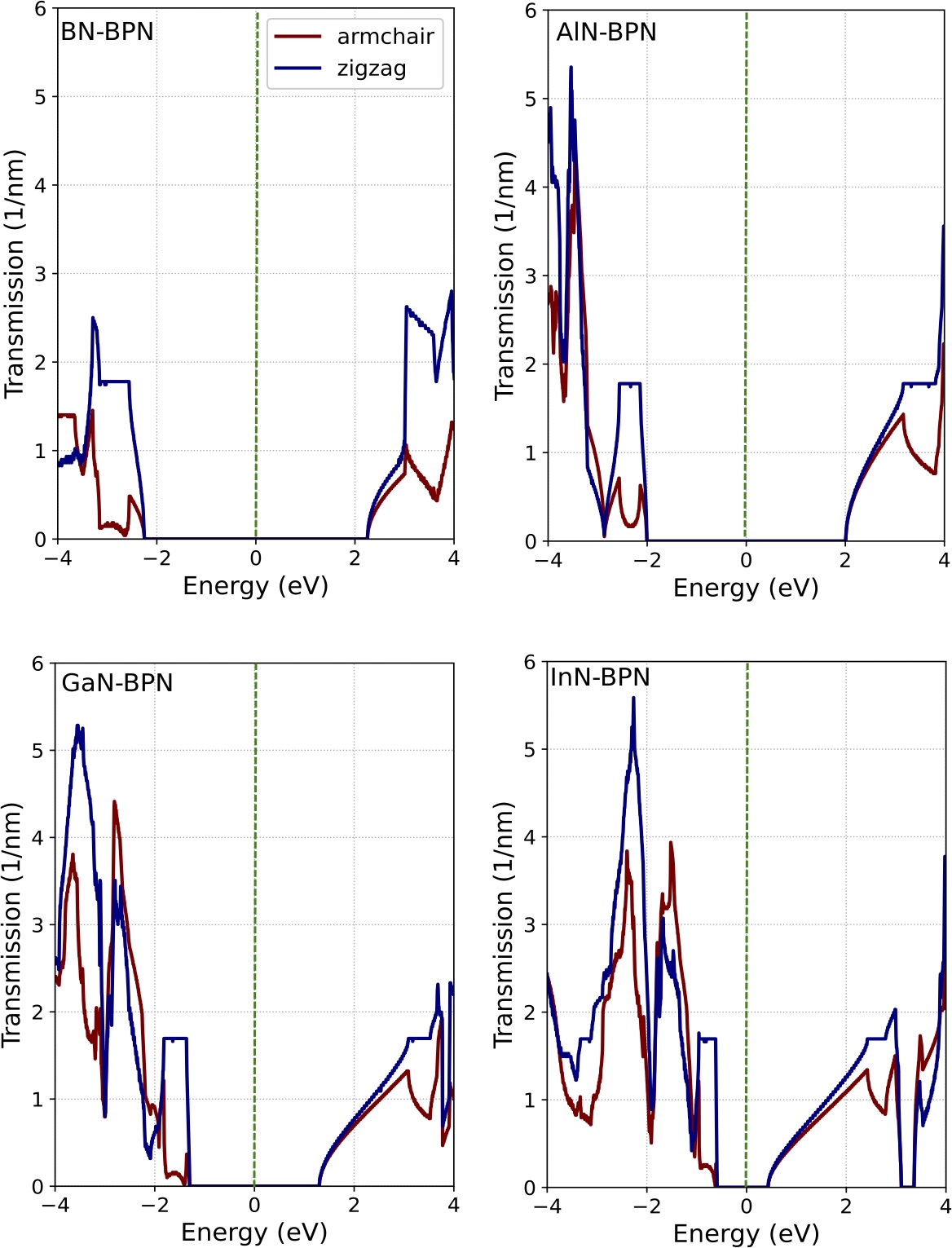}
\caption{\label{fig:fig6}Electronic transmission spectra of group-III nitride biphenylenes along armchair and zigzag directions. Fermi level is set to zero, indicated by green dashed line.}
\end{figure}

\subsubsection{Electronic Transport}

After revealing the structural and electronic properties of group-III nitride BPNs, our focus shifted toward analyzing their electronic transport properties. To make the comparison independent of the specific unit cell size, electronic transmission scaled with width of the unit cell perpendicular to the transport direction. Fig.~\ref{fig:fig6} reflects the strong anisotropy of the $\tau(E)$ along armchair and zigzag directions. In BN-BPN, transmission spectrum changes abruptly near the VB edge along the zigzag direction corresponding to the flat-like band, while it gradually increases at the CB edge for both directions. Although the p-type $\tau(E)$ experiences a sudden increase in armchair direction, as in the zigzag direction, its magnitude is much smaller than in the zigzag direction. For AlN-BPN, $\tau(E)$ gradually adopts a step-like shape in zigzag direction at the VB edge. Likewise, p-type $\tau(E)$ changes more sharply with energy along the armchair direction. For a certain energy range, n-type $\tau(E)$ in both directions coincides and changes with energy more slowly than in the p-type, owing to the nearly quadratic CBM. As moving along the group-III elements, i.e., as for GaN and InN-BPN, variation of the p-type $\tau(E)$ with energy becomes significantly more rapid near the VB edge, while at the CB edge, it gradually adopts a more linear behavior. Changes in the p- and n-type $\tau(E)$ are primarily attributed to the nearly dispersionless region in VB edge and the increasingly dispersive conduction band edge, consistent with the effective mass values presented in Table~\ref{tab:table1}. 

\begin{figure}[htbp]
\includegraphics[width=\linewidth]{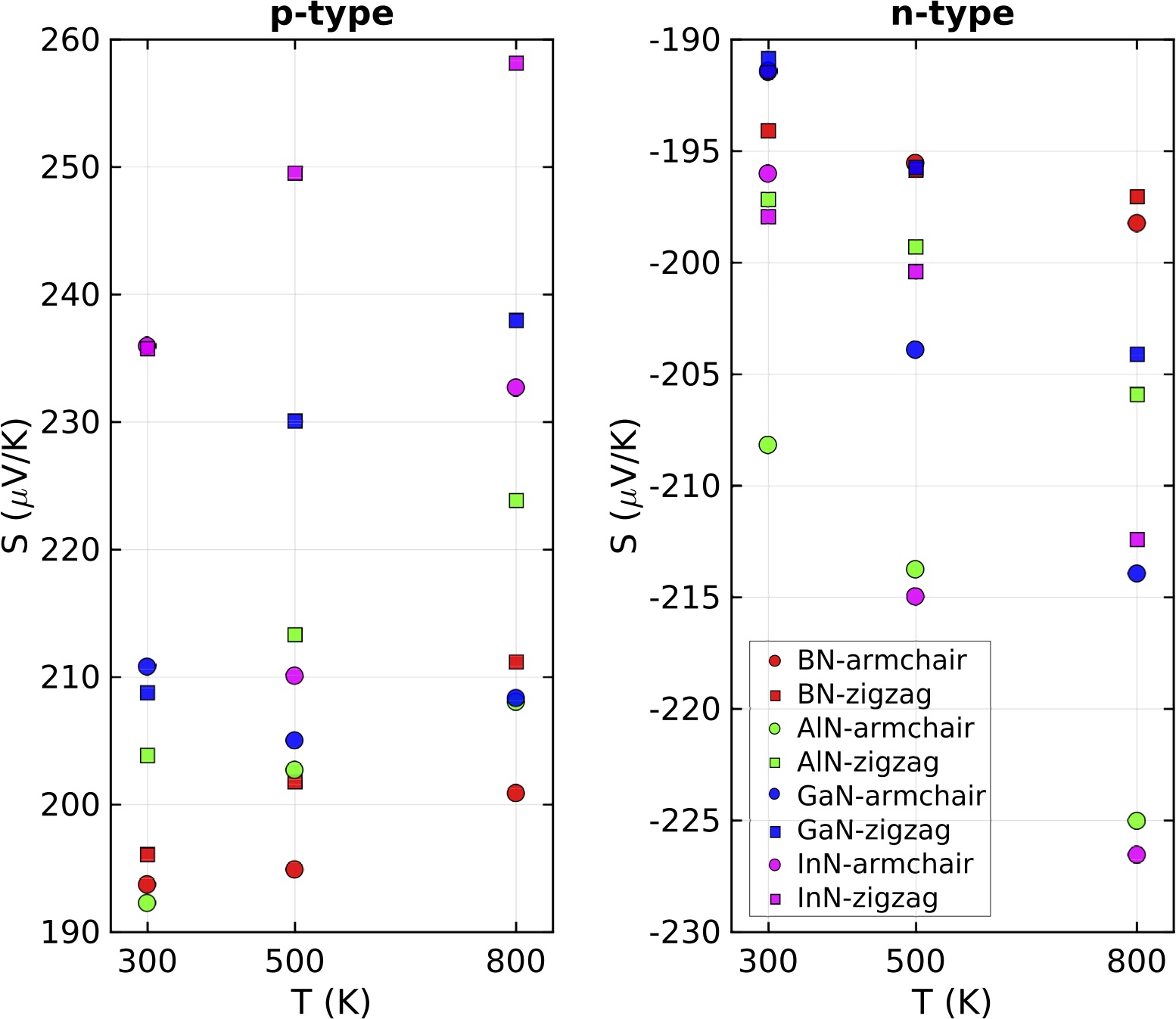}
\caption{\label{fig:fig7}Seebeck coefficient ($S$) values in which $zT$ reaches its maximum value from 300 K to 800 K}
\end{figure}

\begin{figure}[htbp]
\includegraphics[width=\linewidth]{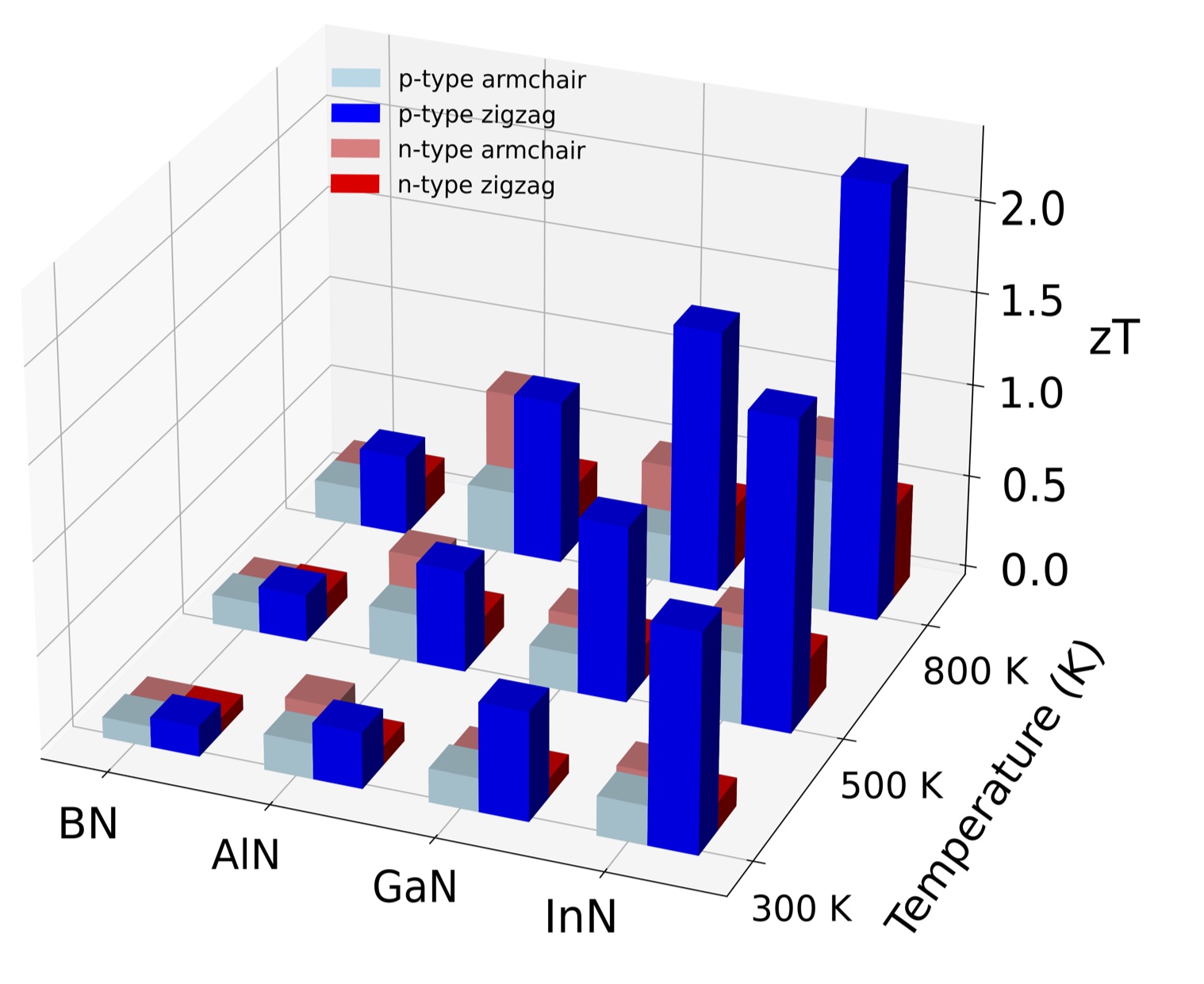}
\caption{\label{fig:fig8}Maximum thermoelectric figure of merit values of group-III nitride biphenylenes along armchair and zigzag directions for various temperatures.}
\end{figure}

After transmission spectra are obtained, one can calculate all the electronic TE coefficients required for evaluating the thermoelectric figure of merit $zT$ (See Fig.S2-S5). The Seebeck coefficients are evaluated at the chemical potential corresponding to the maximum $zT$ is shown in Fig.~\ref{fig:fig7}. At room temperature, InN-BPN possesses the highest p-type $S$ value of 236 $\mu V/K$ for both armchair and zigzag directions thanks to the sudden increase in p-type $\tau(E)$. Although AlN-BPN exhibits a more rapid change in its p-type $\tau(E)$ than BN-BPN along armchair direction, the p-type $S$ attains the lowest $S$ of 192 $\mu V/K$ due to the $\tau(E)$ falling sharply over a narrower energy range. One can observe that, p-type $S$ values of AlN and BN-BPN follow an increasing trend with temperature in both directions, while the $S$ values of GaN and InN-BPN decrease from 300~K to 500~K along the armchair direction. AlN-BPN has the maximum n-type $S$ of -208 $\mu V/K$ at 300~K in armchair direction. As temperature increases, n-type $S$ of all investigated BPNs increases along both directions. At 800~K, n-type $S$ values of AlN and InN-BPN are very close and have the largest value around -225 $\mu V/K$. 
In Fig.~\ref{fig:fig8}, direction-dependent thermoelectric figure of merit $zT$ of group-III nitride BPNs are plotted as a function of temperature for two doping types. For InN-BPN, the combined beneficial effect of the steep increase in $\tau(E)$ and low $\kappa_{ph}$ results in enhanced p-type $zT$ of 1.16 in zigzag direction at room temperature. At 800~K p-type $zT$ of InN-BPN reaches 2.33 along the zigzag direction which is the maximum TE efficiency among all investigated group-III nitride BPNs. 
At room temperature, AlN-BPN exhibits the highest n-type TE performance along the armchair direction, which is accompanied by a remarkably large $S$ despite its high $\kappa_{ph}$. In the armchair direction, the n-type $zT$ value of InN-BPN approaches and marginally exceeds that of AlN-BPN at higher temperatures, reaching 0.80 at 800 K.

\section{CONCLUSION}
In conclusion, first-principles calculations combined with the Landauer formalism are employed to investigate the electronic, thermal, and ballistic thermoelectric properties of group-III nitride biphenylene networks. Phonon dispersion relations an force-field MD simulations were used to asses the stabilities of group-III nitride BPNs  both dynamically and thermally. Electronic band structures calculated using the HSE06 functional show that all considered BPNs are semiconductors with band gaps ranging from 1.02 to 4.51 eV. Pronounced anisotropy is observed in both the electronic and phonon properties, as reflected in the corresponding transmission spectra. As one moves along the group-III elements, the phonon thermal conductance decreases from 1.67 $nW/K/nm$ to 0.25 $nW/K/nm$ and from 1.13 $nW/K/nm$ to 0.13 $nW/K/nm$ in the zigzag and armchair directions, respectively. The flat-like valence-band dispersion between the $\Gamma-X$ symmetry points leads to an abrupt enhancement of the transmission near the valence-band edge, resulting in large p-type Seebeck coefficients in the zigzag direction. By explicitly accounting for the phonon contribution $\kappa_{ph}$, the resulting $zT$ values exhibit a strong dependence on both transport direction and carrier type. In particular, AlN-BPN achieves a maximum n-type $zT$ of 0.80 along the armchair direction, whereas InN-BPN attains the highest p-type $zT$ of 2.33 along the zigzag direction at 800~K. These results provide fundamental insight into electron and phonon transport in group-III nitride biphenylene networks in the ballistic limit and highlight the critical role of anisotropy in optimizing thermoelectric performance.

\begin{acknowledgments}
VOO acknowledges funding from the Scientific and Technological Research Council of Turkey (TUBITAK) under project no 123F264. GOS acknowledges support from the TUBITAK project 123C159. Computational resources have been provided by TUBITAK ULAKBIM, High Performance and Grid Computing Center and by the National Center for High Performance Computing of Turkey (UHeM) under grant number 1013432022.
\end{acknowledgments}



\bibliographystyle{unsrt}

\bibliography{main_manuscript.bib}

\begin{thebibliography}{10}

\bibitem{Geim2007}
A.~K. Geim and K.~S. Novoselov.
\newblock The rise of graphene.
\newblock {\em Nature Materials}, 6(3):183--191, Mar 2007.

\bibitem{doi:10.1126/science.1102896}
K.~S. Novoselov, A.~K. Geim, S.~V. Morozov, D.~Jiang, Y.~Zhang, S.~V. Dubonos,
  I.~V. Grigorieva, and A.~A. Firsov.
\newblock Electric field effect in atomically thin carbon films.
\newblock {\em Science}, 306(5696):666--669, 2004.

\bibitem{PhysRevLett.106.155703}
Xian-Lei Sheng, Qing-Bo Yan, Fei Ye, Qing-Rong Zheng, and Gang Su.
\newblock T-carbon: A novel carbon allotrope.
\newblock {\em Phys. Rev. Lett.}, 106:155703, Apr 2011.

\bibitem{JIANG2017370}
Jin-Wu Jiang, Jiantao Leng, Jianxin Li, Zhengrong Guo, Tienchong Chang,
  Xingming Guo, and Tongyi Zhang.
\newblock Twin graphene: A novel two-dimensional semiconducting carbon
  allotrope.
\newblock {\em Carbon}, 118:370--375, 2017.

\bibitem{C8RA07074A}
Li~Zeng, Yingxiang Cai, Zhihao Xiang, Yu~Zhang, and Xuechun Xu.
\newblock A new metallic $\pi$-conjugated carbon sheet used for the cathode of
  li–s batteries.
\newblock {\em RSC Adv.}, 9:92--98, 2019.

\bibitem{LIMA2025117235}
Kleuton~A.L. Lima, José~A.S. Laranjeira, Nicolas~F. Martins, Alexandre~C.
  Dias, Julio~R. Sambrano, Douglas~S. Galvão, and Luiz A.~Ribeiro Junior.
\newblock Petal-graphyne: A novel 2d carbon allotrope for high-performance li
  and na ion storage.
\newblock {\em Journal of Energy Storage}, 130:117235, 2025.

\bibitem{doi:10.1073/pnas.1416591112}
Shunhong Zhang, Jian Zhou, Qian Wang, Xiaoshuang Chen, Yoshiyuki Kawazoe, and
  Puru Jena.
\newblock Penta-graphene: A new carbon allotrope.
\newblock {\em Proceedings of the National Academy of Sciences},
  112(8):2372--2377, 2015.

\bibitem{LARANJEIRA2024100321}
José~A.S. Laranjeira, Nicolas~F. Martins, Pablo~A. Denis, and Julio~R.
  Sambrano.
\newblock Graphenyldiene: A new sp2-graphene-like nanosheet.
\newblock {\em Carbon Trends}, 14:100321, 2024.

\bibitem{Du2017}
Qi-Shi Du, Pei-Duo Tang, Hua-Lin Huang, Fang-Li Du, Kai Huang, Neng-Zhong Xie,
  Si-Yu Long, Yan-Ming Li, Jie-Shan Qiu, and Ri-Bo Huang.
\newblock A new type of two-dimensional carbon crystal prepared from
  1,3,5-trihydroxybenzene.
\newblock {\em Scientific Reports}, 7(1):40796, Jan 2017.

\bibitem{C2TC00006G}
Qi~Song, Bing Wang, Ke~Deng, Xinliang Feng, Manfred Wagner, Julian~D. Gale,
  Klaus Müllen, and Linjie Zhi.
\newblock Graphenylene{,} a unique two-dimensional carbon network with
  nondelocalized cyclohexatriene units.
\newblock {\em J. Mater. Chem. C}, 1:38--41, 2013.

\bibitem{GIRAO2023611}
Eduardo~Costa Girão, Alastair Macmillan, and Vincent Meunier.
\newblock Classification of sp2-bonded carbon allotropes in two dimensions.
\newblock {\em Carbon}, 203:611--619, 2023.

\bibitem{tyutyulkov1997structure}
Nikolai Tyutyulkov, F~Dietz, Klaus M{\"u}llen, and Martin Baumgarten.
\newblock Structure and energy spectra of a two-dimensional dielectric carbon
  allotrope.
\newblock {\em Chem. Phys. Lett.}, 272(1-2):111--114, 1997.

\bibitem{hudspeth2010electronic}
Mathew~A Hudspeth, Brandon~W Whitman, Veronica Barone, and Juan~E Peralta.
\newblock Electronic properties of the biphenylene sheet and its
  one-dimensional derivatives.
\newblock {\em ACS nano}, 4(8):4565--4570, 2010.

\bibitem{karaush2014dft}
Nataliya~N Karaush, Gleb~V Baryshnikov, and Boris~F Minaev.
\newblock Dft characterization of a new possible graphene allotrope.
\newblock {\em Chem. Phys. Lett.}, 612:229--233, 2014.

\bibitem{doi:10.1126/science.abg4509}
Qitang Fan, Linghao Yan, Matthias~W. Tripp, Ondřej Krejčí, Stavrina
  Dimosthenous, Stefan~R. Kachel, Mengyi Chen, Adam~S. Foster, Ulrich Koert,
  Peter Liljeroth, and J.~Michael Gottfried.
\newblock Biphenylene network: A nonbenzenoid carbon allotrope.
\newblock {\em Science}, 372(6544):852--856, 2021.

\bibitem{https://doi.org/10.1002/anie.201707272}
Ann-Kristin Steiner and Konstantin~Y. Amsharov.
\newblock The rolling-up of oligophenylenes to nanographenes by a hf-zipping
  approach.
\newblock {\em Angewandte Chemie International Edition}, 56(46):14732--14736,
  2017.

\bibitem{D2NH00528J}
Kaifeng Niu, Qitang Fan, Lifeng Chi, Johanna Rosen, J.~Michael Gottfried, and
  Jonas Björk.
\newblock Unveiling the formation mechanism of the biphenylene network.
\newblock {\em Nanoscale Horiz.}, 8:368--376, 2023.

\bibitem{demirci2022hydrogenated}
Salih Demirci, Taylan Gorkan, Şafak Çall{\i}oğlu, V.~Ongun Ozcelik,
  Johannes~V Barth, Ethem Aktürk, and Salim Ciraci.
\newblock Hydrogenated carbon monolayer in biphenylene network offers a
  potential paradigm for nanoelectronic devices.
\newblock {\em The Journal of Physical Chemistry C}, 126(36):15491--15500,
  2022.

\bibitem{demirci2022stability}
Salih Demirci, {\c{S}}afak {\c{C}}all{\i}o{\u{g}}lu, Taylan G{\"o}rkan, Ethem
  Akt{\"u}rk, and Salim Ciraci.
\newblock Stability and electronic properties of monolayer and multilayer
  structures of group-iv elements and compounds of complementary groups in
  biphenylene network.
\newblock {\em Physical Review B}, 105(3):035408, 2022.

\bibitem{OZDEMIR2025118280}
Ilkay Ozdemir, Eren~Ege Karacan, Ethem Aktürk, Gökhan Gökoglu, Johannes~V.
  Barth, and Olcay Üzengi Aktürk.
\newblock Alloy-driven modulation of electronic, mechanical and optical
  properties in 2d bpn-ga1-xalxn monolayers.
\newblock {\em Materials Science and Engineering: B}, 319:118280, 2025.

\bibitem{abdullahi2023stability}
Yusuf~Zuntu Abdullahi and Fatih Ersan.
\newblock Stability and electronic properties of xo (x: Be, mg, zn, cd)
  biphenylene and graphenylene networks: A first-principles study.
\newblock {\em Applied Physics Letters}, 123(25), 2023.

\bibitem{laranjeira2024zns}
Jos{\'e}~AS Laranjeira, Yusuf~Z Abdullahi, Fatih Ersan, and Julio~R Sambrano.
\newblock Zns and cds counterparts of biphenylene lattice: A density functional
  theory prediction.
\newblock {\em Computational and Theoretical Chemistry}, 1235:114580, 2024.

\bibitem{gorkan2023can}
Taylan Gorkan, Salih Demirci, Johannes~V Barth, Ethem Aktürk, and Salim
  Ciraci.
\newblock Can stable mos2 monolayers and multilayers be constituted in the
  biphenylene network?
\newblock {\em The Journal of Physical Chemistry C}, 127(18):8770--8777, 2023.

\bibitem{gorkan2022functional}
Taylan Gorkan, Şafak Çall{\i}oğlu, Salih Demirci, Ethem Aktürk, and Salim
  Ciraci.
\newblock Functional carbon and silicon monolayers in biphenylene network.
\newblock {\em ACS Applied Electronic Materials}, 4(6):3056--3070, 2022.

\bibitem{D3CP00776F}
Mukesh Singh and Brahmananda Chakraborty.
\newblock 2d bn-biphenylene: structure stability and properties tenability from
  a dft perspective.
\newblock {\em Phys. Chem. Chem. Phys.}, 25:16018--16029, 2023.

\bibitem{Santos2023}
Emanuel J.~A. Santos, William~F. Giozza, Rafael~T. de~Souza~J{\'u}nior,
  Neymar~J. Nepomuceno~Cavalcante, Luiz~A. Ribeiro~J{\'u}nior, and Kleuton~A.
  Lopes~Lima.
\newblock On the co{\$}{\$}{\_}{\{}2{\}}{\$}{\$}adsorption in a boron nitride
  analog for the recently synthesized biphenylene network: a dft study.
\newblock {\em Journal of Molecular Modeling}, 29(10):327, Sep 2023.

\bibitem{MOHAMMED2025112535}
Khidhair~Jasim Mohammed, Abdulrahman~T. Ahmed, Mandeep Kaur, Prakash Kanjariya,
  Junainah~Abd Hamid, Bharti Kumari, Mukhlisa Soliyeva, Suman Saini, and
  Abdulrahman~A. Almehizia.
\newblock Rational design of two-dimensional bn biphenylene monolayer as an
  efficient anode material for ca-ion batteries: A theoretical study.
\newblock {\em Journal of Physics and Chemistry of Solids}, 199:112535, 2025.

\bibitem{Jardan2025}
Yousef A.~Bin Jardan, Yashwantsinh Jadeja, Soumya~V. Menon, Debasish Shit,
  Girish~Chandra Sharma, Mounir~M. Bekhit, Rasha~Ali Abdalhuseen, Salah
  Hassan~Zain Al-Abdeen, and Ruaa Sattar.
\newblock Utilizing bn-biphenylene nanosheet as a potential carrier for
  thioguanine: a promising drug delivery system.
\newblock {\em Chemical Papers}, 79(4):2531--2540, Apr 2025.

\bibitem{D4NR02754J}
Ajay Kumar, Parbati Senapati, and Prakash Parida.
\newblock Unravelling the potential of a hybrid borocarbonitride biphenylene 2d
  network for thermoelectric applications: a first principles study.
\newblock {\em Nanoscale}, 17:4015--4029, 2025.

\bibitem{D4NR01386G}
Ajay Kumar and Prakash Parida.
\newblock Unveiling the potential of a bcn-biphenylene monolayer as a
  high-performance anode material for alkali metal ion batteries: a
  first-principles study.
\newblock {\em Nanoscale}, 16:13131--13147, 2024.

\bibitem{YUAN2024159555}
Yuan Yuan, Jin {Yong Lee}, Shaul Mukamel, and Baotao Kang.
\newblock Strain engineering enhances electrocatalytic hydrogen evolution of
  b-n pair substituted biphenylene networks.
\newblock {\em Applied Surface Science}, 655:159555, 2024.

\bibitem{doi:10.1021/acsomega.4c03511}
Jos{\'e} A.~S. Laranjeira, Nicolas Martins, Pablo~A. Denis, and Julio Sambrano.
\newblock Unveiling a new 2d semiconductor: Biphenylene-based inn.
\newblock {\em ACS Omega}, 9(26):28879--28887, 2024.

\bibitem{LOPESLIMA2023107183}
K.A. {Lopes Lima} and L.A. Ribeiro.
\newblock A dft study on the mechanical, electronic, thermodynamic, and optical
  properties of gan and aln counterparts of biphenylene network.
\newblock {\em Materials Today Communications}, 37:107183, 2023.

\bibitem{ABDULLAHI2024113103}
Yusuf~Zuntu Abdullahi and Fatih Ersan.
\newblock New stable inorganic bx (x= p, as, sb) biphenylene and graphenylene
  monolayers: A first-principles investigation.
\newblock {\em Computational Materials Science}, 242:113103, 2024.

\bibitem{DJEBABLIA202433}
Ikram Djebablia, Yusuf~Zuntu Abdullahi, Kamel Zanat, and Fatih Ersan.
\newblock Metal-decorated boron phosphide (bp) biphenylene and graphenylene
  networks for ultrahigh hydrogen storage.
\newblock {\em International Journal of Hydrogen Energy}, 66:33--39, 2024.

\bibitem{doi:10.1021/acsaem.5c00897}
Divya Dange, Preeti Beniwal, and Thogluva~Janardhanan Dhilip~Kumar.
\newblock Efficient hydrogen storage in superalkali nli4-decorated boron
  phosphide biphenylene.
\newblock {\em ACS Applied Energy Materials}, 8(11):7594--7604, 2025.

\bibitem{D4RA07271E}
Yusuf~Zuntu Abdullahi, Ikram Djebablia, Tiem~Leong Yoon, and Lim~Thong Leng.
\newblock Calcium-atom-modified boron phosphide (bp) biphenylene as an
  efficient hydrogen storage material.
\newblock {\em RSC Adv.}, 14:39268--39275, 2024.

\bibitem{HAOUAM2024160096}
Marwa Haouam, Yusuf~Zuntu Abdullahi, Kamel Zanat, and Fatih Ersan.
\newblock Boron phosphide (bp) biphenylene and graphenylene networks as anode
  and anchoring materials for li/na-ion and li/na–s batteries.
\newblock {\em Applied Surface Science}, 662:160096, 2024.

\bibitem{ZAINUL2024112805}
Rahadian Zainul, Ali Basem, Ahmed~K. Nemah, Raman Kumar, Rakesh Kumar, Rohit
  Sharma, Aseel~M. Aljeboree, Mamata Chahar, and Yasser Elmasry.
\newblock Bp-biphenylene monolayers: A promising anode material for
  high-performance magnesium-ion batteries.
\newblock {\em Inorganic Chemistry Communications}, 168:112805, 2024.

\bibitem{hicks1993}
L.~D. Hicks and M.~S. Dresselhaus.
\newblock Effect of quantum-well structures on the thermoelectric figure of
  merit.
\newblock {\em Phys. Rev. B}, 47:12727--12731, May 1993.

\bibitem{PhysRevMaterials.9.024003}
G\"ozde~\"Ozbal Sargin, Salih Demirci, Kai Gong, and V.~Ongun
  \"Oz\ifmmode~\mbox{\c{c}}\else \c{c}\fi{}elik.
\newblock Ultrathin carbon biphenylene network as an anisotropic thermoelectric
  material with high temperature stability under mechanical strain.
\newblock {\em Phys. Rev. Mater.}, 9:024003, Feb 2025.

\bibitem{PhysRevB.107.045422}
Fang Lv, Hanpu Liang, and Yifeng Duan.
\newblock Funnel-shaped electronic structure and enhanced thermoelectric
  performance in ultralight
  ${\mathrm{c}}_{x}{(\mathrm{BN})}_{1\text{\ensuremath{-}}x}$ biphenylene
  networks.
\newblock {\em Phys. Rev. B}, 107:045422, Jan 2023.

\bibitem{D3CP03088A}
Ajay Kumar, Parbati Senapati, and Prakash Parida.
\newblock Theoretical insights into the structural{,} electronic and
  thermoelectric properties of the inorganic biphenylene monolayer.
\newblock {\em Phys. Chem. Chem. Phys.}, 26:2044--2057, 2024.

\bibitem{hafner1997vienna}
J{\"u}rgen Hafner and Georg Kresse.
\newblock The vienna ab-initio simulation program vasp: An efficient and
  versatile tool for studying the structural, dynamic, and electronic
  properties of materials.
\newblock In {\em Properties of Complex Inorganic Solids}, pages 69--82.
  Springer, 1997.

\bibitem{kresse1996efficient}
Georg Kresse and J{\"u}rgen Furthm{\"u}ller.
\newblock Efficient iterative schemes for ab initio total-energy calculations
  using a plane-wave basis set.
\newblock {\em Phys. Rev. B}, 54(16):11169, 1996.

\bibitem{Smidstrup_2020}
Søren Smidstrup, Troels Markussen, Pieter Vancraeyveld, Jess Wellendorff,
  Julian Schneider, Tue Gunst, Brecht Verstichel, Daniele Stradi, Petr~A
  Khomyakov, Ulrik~G Vej-Hansen, Maeng-Eun Lee, Samuel~T Chill, Filip
  Rasmussen, Gabriele Penazzi, Fabiano Corsetti, Ari Ojanperä, Kristian
  Jensen, Mattias L~N Palsgaard, Umberto Martinez, Anders Blom, Mads Brandbyge,
  and Kurt Stokbro.
\newblock Quantumatk: an integrated platform of electronic and atomic-scale
  modelling tools.
\newblock {\em Journal of Physics: Condensed Matter}, 32(1):015901, oct 2019.

\bibitem{VANSETTEN201839}
M.J. {van Setten}, M.~Giantomassi, E.~Bousquet, M.J. Verstraete, D.R. Hamann,
  X.~Gonze, and G.-M. Rignanese.
\newblock The pseudodojo: Training and grading a 85 element optimized
  norm-conserving pseudopotential table.
\newblock {\em Computer Physics Communications}, 226:39--54, 2018.

\bibitem{zhou2013molecular}
Xiao~Wang Zhou, Reese~E Jones, John~C Duda, and Patrick~E Hopkins.
\newblock Molecular dynamics studies of material property effects on thermal
  boundary conductance.
\newblock {\em Phys. Chem. Chem. Phys.}, 15(26):11078--11087, 2013.

\bibitem{moon2007theoretical}
Won~Ha Moon, Myung~Sik Son, and Ho~Jung Hwang.
\newblock Theoretical study on structure of boron nitride fullerenes.
\newblock {\em Applied Surface Science}, 253(17):7078--7081, 2007.

\bibitem{zhou2015towards}
Xiaowang Zhou and Reese~E Jones.
\newblock Towards molecular dynamics simulations of ingan nanostructures.
\newblock Technical report, Sandia National Lab.(SNL-CA), Livermore, CA (United
  States), 2015.

\bibitem{martyna1992nose}
Glenn~J Martyna, Michael~L Klein, and Mark Tuckerman.
\newblock Nos{\'e}--hoover chains: The canonical ensemble via continuous
  dynamics.
\newblock {\em The Journal of chemical physics}, 97(4):2635--2643, 1992.

\bibitem{10.1063/1.3531573}
Jin-Wu Jiang, Jian-Sheng Wang, and Baowen Li.
\newblock A nonequilibrium green’s function study of thermoelectric
  properties in single-walled carbon nanotubes.
\newblock {\em Journal of Applied Physics}, 109(1):014326, 01 2011.

\bibitem{sivan1986multichannel}
U~Sivan and Y~Imry.
\newblock Multichannel landauer formula for thermoelectric transport with
  application to thermopower near the mobility edge.
\newblock {\em Physical review b}, 33(1):551, 1986.

\bibitem{esfarjani2006thermoelectric}
Keivan Esfarjani, Mona Zebarjadi, and Yoshiyuki Kawazoe.
\newblock Thermoelectric properties of a nanocontact made of two-capped
  single-wall carbon nanotubes calculated within the tight-binding
  approximation.
\newblock {\em Physical Review B}, 73(8):085406, 2006.

\bibitem{datta1997electronic}
Supriyo Datta.
\newblock {\em Electronic transport in mesoscopic systems}.
\newblock Cambridge university press, 1997.

\bibitem{rego1998quantized}
Luis~GC Rego and George Kirczenow.
\newblock Quantized thermal conductance of dielectric quantum wires.
\newblock {\em Physical Review Letters}, 81(1):232, 1998.

\bibitem{sevinccli2019green}
Haldun Sevin{\c{c}}li, Stephan Roche, Gianaurelio Cuniberti, Mads Brandbyge,
  Rafael Guti{\'e}rrez, and L~Medrano Sandonas.
\newblock Green function, quasi-classical langevin and kubo--greenwood methods
  in quantum thermal transport.
\newblock {\em Journal of Physics: Condensed Matter}, 31(27):273003, 2019.

\bibitem{data}
{The data are available upon reasonable request from the authors. }.

\end{thebibliography}

\end{document}